# Title: Decoupled phase modulation for circularly polarized lights via chiral metasurface

Renchao Jin,[1] Lin Deng,[3] Lili Tang,[1] Yue Cao,[1] Yongmin Liu,[2,3,#] and Zheng-Gao Dong[1,*]

[1]School of Physics, Southeast University, Nanjing 211189, China

[2]Department of Mechanical and Industrial Engineering, Northeastern University, Boston, Massachusetts 02115, USA

[3]Department of Electrical and Computer Engineering, Northeastern University, Boston, Massachusetts 02115, USA

**Abstract**:

Metasurfaces are believed as one of the best candidates in nano-optical devices, attributed to the key feasible modulation features of phase, polarization, and local field enhancement by structural designing. However, current methods of propagation- and geometric-phase modulation are interrelated between two eigen spin-states. This means that when the left-handed component phase of a beam is modulated by metasurfaces, its right-handed component phase will change accordingly, which limits the versatility of spin-decoupled applications. In this paper, we experimentally and numerically demonstrate a new phase modulation pathway based on chiral V-shaped holes, which enable fully decoupled one-handed phase modulation of the two eigen spin-states. Two enantiomers are proposed to realize decoupled functions for the two eign-states, e.g., the enantiomer can manipulate the left-handed component phase of a laser beam without changes of the right-handed component. This proposed method has significant meaning in metasurfaces, which can expand the methods of phase engineering.

**Keywords:** Metasurfaces, photonic spin Hall effect, plasmonic

**Introduction:**

The manipulation of the electromagnetic wave is one of the vital things for the development of modern optoelectronic technology. Metasurface, as an artificial metamaterial with a subwavelength thickness that enables us to flexibly manipulate electromagnetic waves, has been widely studied [1–3]. Subwavelength structures that constitute the metasurface can modulate the amplitude, phase, and polarization of electromagnetic waves. Among these properties, phase modulation is of much interest, as they have already been used to generate beam steering [4–6], hologram [7–9], focusing [10–12], and so on. At first, the metasurface could selectively control the phase of a particular linearly polarized light [1], which was easily achieved by changing geometric parameters of subwavelength structures. Thereafter, a single metasurface with multiple phase modulations for different linearly polarized lights (i.e., transverse magnetic and transverse electric modes) was investigated for a variety of applications [13]. Recently, the phase modulation of left-handed and right-handed circularly polarized lights (i.e., LCP and RCP, respectively) attracted much attention, since they as two orthogonal eigen spin states open up many applications while interacting with matters, such as chiral sensing [14,15], polarization imaging [16,17], information coding [18–20], and spin-orbital interaction [21–23].

The conventional methods to manipulate the phase of LCP and RCP lights are divided into so-called propagation-phase and geometric-phase (Pancharatnam-Berry, PB-phase) modulations [24–32]. The former is related to the geometric parameters of subwavelength structures and always imparts the same phase modulation to LCP and RCP lights simultaneously [24]. Due to its identical phase modulation for LCP and RCP incidences, the propagation phase is a good candidate for circular-polarization insensitive devices [33,34]. In contrast, the geometric phase imparts different phase modulation to LCP and RCP lights, with opposite signs but the same absolute value. Therefore, geometric-phase modulation is used to split LCP and RCP lights, for example, tremendous devices were fabricated to study the photonic spin Hall effect [27,35,36]. However, both propagation-phase and geometric-phase modulation methods have their limits, and we can hardly achieve decoupled phase modulation for LCP and RCP lights by either of methods (i.e., modulating LCP phase while leaving RCP phase unmodulated, or vice versa), though for linear polarized light it can be easily achieved [13]. Fortunately, by combining propagation and geometric phases, the arbitrary phase modulation for LCP and RCP lights has been achieved, which has expanded the fields in spin-dependent optical devices [24–28]. Even though the method can manipulate the phase of LCP and RCP lights simultaneously, the phase modulation is still not decoupled (e.g., while the phase of LCP light is modulated, the phase of RCP light will be affected inevitably). The development in metasurfaces asks a question, can we decouple the phase modulation for these two spin states of light?

In this letter, we demonstrate a new approach to decouple the phase for LCP and RCP light. Based on chiral metasurfaces, which can uniquely manipulate the phase of either LCP or RCP without geometric phase involved. The chiral metasurfaces have fundamental constituted structures which are schematically plotted in **Figure 1a-c**. We designed a V-shaped hole that was etched in a 40-nm-thickness silver film. It has two tunable arms, i.e., the top arm $L_{top}$ and bottom arm $L_{bot}$, and an identical width W = 100 nm, respectively. The V-shaped holes plotted in **Figure 1a-c** contain two enantiomers and symmetric one, where the V-shaped hole with tunable bottom arm $L_{bot}$ (the top arm $L_{top}$ is set to 430 nm) is enantiomer A (Ent A), as shown in **Figure 1a**. Then, in **Figure 1b**, the symmetric V-shaped hole has two tunable arms of the same length, called symmetry (Sym). The Enantiomer B (Ent B) in **Figure 1c** has the opposite chirality with an adjustable top arm $L_{top}$. Depending on the configurations, these holes should have different phase responses. Therefore, we studied their phase responses for LCP and RCP incidences via the commercial software (CST Studio Suite), the simulated results are shown in **Figure 1d-f**. In the following, the LCP and RCP lights are symbolized as – and + respectively, for example, the symbol $\angle t_{+-}$ represents the phase of transmitted RCP component with LCP incidence. According to the results, we find that Ent A, in contrast to Ent B and Sym, has a large phase response to the change of $L_{bot}$ when the incidence is RCP (solid orange line in **Figure 1d**). As for its counterpart, the rapidly changed solid blue line in **Figure 1f** indicates that Ent B is sensitive to LCP incidence. Then the results in **Figure 1e** for nonchiral Sym show identical phase responses to LCP and RCP incidences. Therefore, the simulated results here demonstrate the ability to decouple the phase responses of LCP and RCP light, using Ent A and Ent B. We want to emphasize that the enantiomers have no obvious circular dichroism [30,37–41], since the amplitudes of transmission for LCP and RCP incidences almost keep the same (supporting information, **Figure S1**). In other words, by using Ent A and Ent B, we can readily decouple the phase of LCP and RCP, this has important applications in spin-decoupled metasurface design.

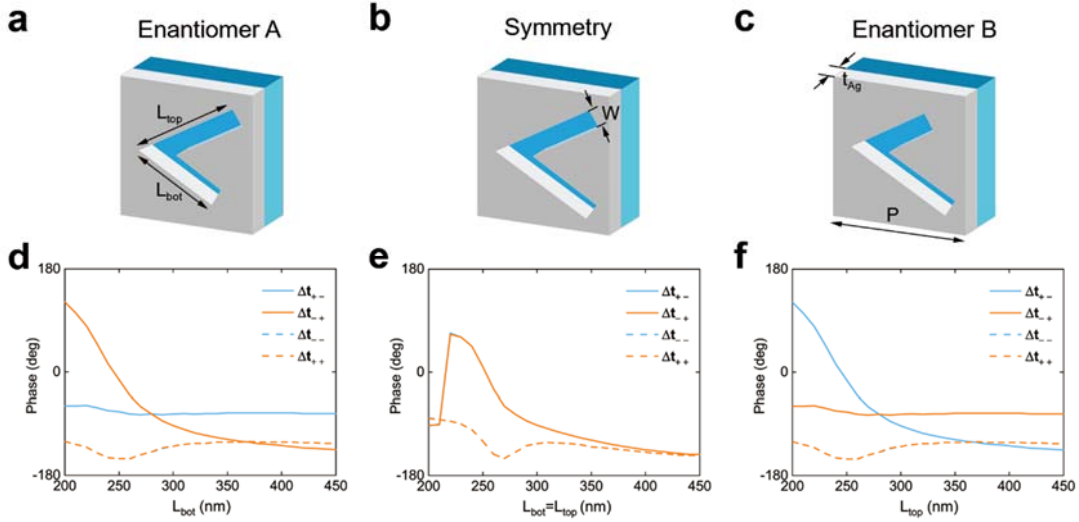

**Figure 1.** Schematic of (a) enantiomer A (b) symmetry and (c) enantiomer B, respectively. The corresponding phase responses of enantiomer A (d), symmetry (e), and enantiomer B (f). The geometric parameters are W = 100 nm, P = 450 nm, and $t_{Ag}$ = 40 nm.

To explicitly demonstrate the decoupled phase modulation of the enantiomers, we simulated the phase diagrams of orthogonal CP conversions at 1064 nm for RCP and LCP incidences, the results are depicted in **Figure 2a** and **2b**, respectively. The phase modulation of RCP-to-LCP conversion in **Figure 2a** shows different phase values along the horizontal direction but the same values along the vertical direction, indicating only $L_{bot}$ can manipulate the phase for RCP-to-LCP conversion. While for LCP-to-RCP conversion, **Figure 2b** shows a mirror symmetry of **Figure 2a**, indicating only $L_{top}$ can manipulate the phase for LCP-to-RCP conversion. These results indicate by designing some specific enantiomers, we can achieve fully decoupled phase modulation for the two orthogonal CP conversions. This enantiomers-design strategy is simpler than the existing method combing propagation phase and geometric phase to achieve fully spin-decoupled phase modulation [24-32]. Because the enantiomers do not involve in-plane rotations, the geometric phase is absent in our work, which can also reduce the difficulty in metasurface fabrications.

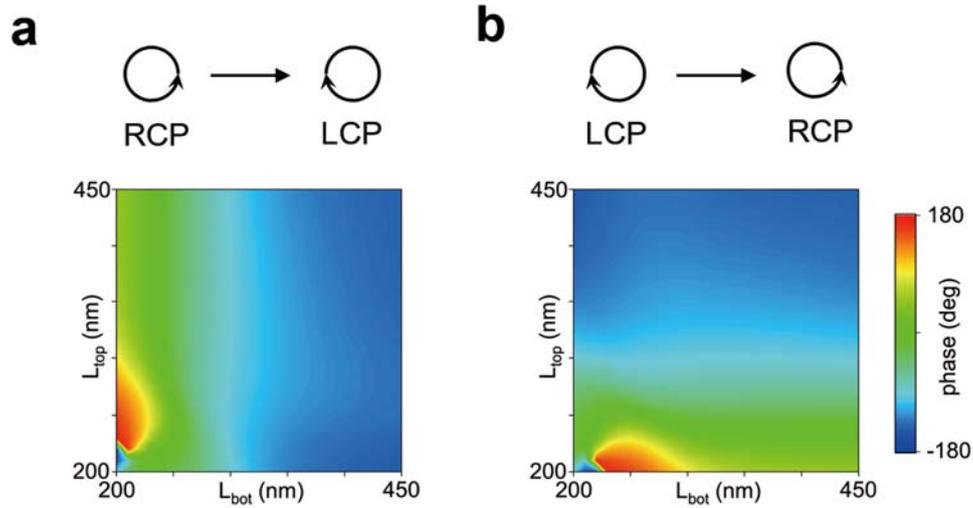

**Figure 2.** Phase diagrams for the CP conversions, (a) RCP-to-LCP conversion, (b) LCP-to-RCP conversion (Counterclockwise for RCP, clockwise for LCP ).

Then, to clarify the reason for spin-decoupled phase modulation in our enantiomers, we also simulated the near magnetic field distributions of enantiomers and symmetry, the results are shown in **Figure 3a-f**. For the Sym, the magnetic field shows chiral distribution depending on RCP and LCP incidences, where the magnetic fields are more concentrated in the bottom/top arm for RCP/LCP incidence (**Figure 3b/3e**). Therefore, when we decrease the length of the bottom arm for Ent A, the magnetic field distribution changes a lot for RCP incidence, resulting in a larger local enhancement in the bottom arm (**Figure 3d**). However, when input with LCP, although the length of the bottom arm decreases, the magnetic field distribution (**Figure 3a**) almost keeps the same as Sym one (**Figure 3b**). On the contrary, Ent B shows a concentrated magnetic field in the top arm while illuminated with LCP. Consequently, due to these distinguished near magnetic field distributions on enantiomers for RCP and LCP incidences, the spin-decoupled phase modulation can be achieved by applying independent enantiomer manipulations, in contrast to the geometric-phase modulation with opposite rather than arbitrary CP-conversion phases. The chiral field distribution presented in symmetric V-shaped antennas has been well studied in the literature [42,43], which was attributed to the interference of resonant modes in the V-shaped antenna.

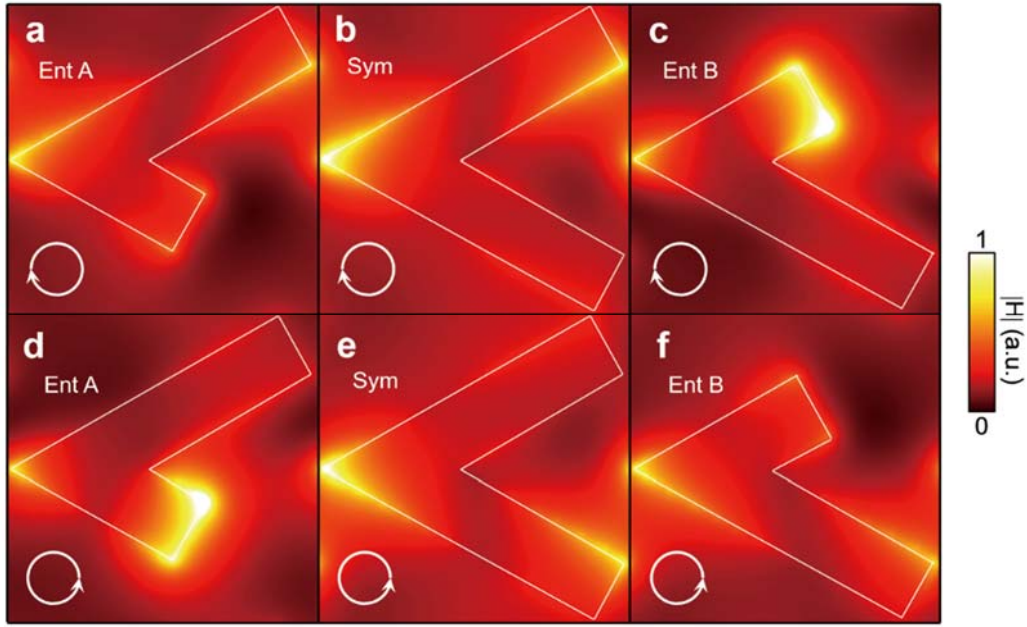

**Figure 3.** The near magnetic field intensity of Ent A (a), Sym (b), and Ent B (c) for LCP incidence. The near magnetic field intensity of Ent A (d), Sym (e), and Ent B (f) for RCP incidence. The long arm is 430 nm, while the short arm is 260 nm.

The decoupled phase modulation by using enantiomers reported here is completely different from existing propagation-phase and geometric-phase modulation methods. As a proof-of-principle demonstration, an Ent A metasurface was fabricated to confirm our theoretical analysis. The metasurface in **Figure 4a** consisted of four Ent A to form a phase gradient of $2\pi/4 \cdot P$ for RCP incidence (from left to right in one supercell, the $L_{bot}$ varied as 430 nm, 260 nm, 200 nm, and 260 nm). This means that it only deflects RCP component of incident beam. Note, the fourth Ent A has a rotation angle of 90 degrees, because the Ent A only reach a phase modulation ranging from 0 to 250 degrees in this case, but the complementary structures (V-shaped antenna on the metallic substrate) can reach a full phase modulation ranging from 0 to 360 degrees based on our simulation [supporting information, **Figure S2** and **S3**]. Then, we experimentally characterized the metasurface using a homemade optical path. A continuous laser centered at 1064 nm was used as the light source. A pair of half-wave plate and quarter-wave plate was placed in front of the laser to generate LCP and RCP incidences, and another pair of quarter-wave plate and linear polarizer was placed behind the sample to detect the polarization of transmitted light (supporting information,

**Figure S4**). In the optical path, we designed a quasi-collimated illumination by the pair of an objective lens and a plano-convex lens. Therefore, the incident beam has an invariant diameter of around 5 μm after passing through the sample. The measured transmitted powers are shown in **Figure 4b** and **4c**. Obviously, for RCP incidence, the detected LCP component (CP conversion component) of transmitted light has an obvious deflection due to the phase modulation of Ent A. The angle of refraction in **Figure 4b** is 35.75 degrees, which is in perfect agreement with our full-wave simulation (36.2 degrees in **Figure 4d**). However, if we flip the polarization of the input laser beam to LCP, the detected RCP component of the transmitted light shows a refraction angle of 0 degree, which is as expected from the full-wave simulation in **Figure 4e**. Furthermore, the measured co-polarized components have a refraction angle of 0 degree, since our spin-decoupled phase modulation is based on CP conversions (see supporting information, **Figure S5**). As for the experimental efficiency, the transmissivity in **Figure 4b** is measured as 7.6%, while the transmissivity in **Figure 4c** is 9.3%, and the slightly lower efficiency of the former is due to sample imperfection. Although the efficiency of the reported metallic metasurface here is not that high, it can be further enhanced by using either dielectric metasurface or reflection-type metasurface, where the decoupled phase modulation also exists. Chiral metasurfaces composed of Ent B were also fabricated and measured (Supporting Information, **Figures S6** and **S7**), and only LCP incidence can cause anomalous refraction as expected. In addition, for a comprehensive comparison, Sym-based metasurfaces were fabricated and measured, and the results are shown in **Figures S8** and **S9**. It is obvious that the refraction angles of LCP and RCP are the same, in other words, the spin-phase response is not decoupled.

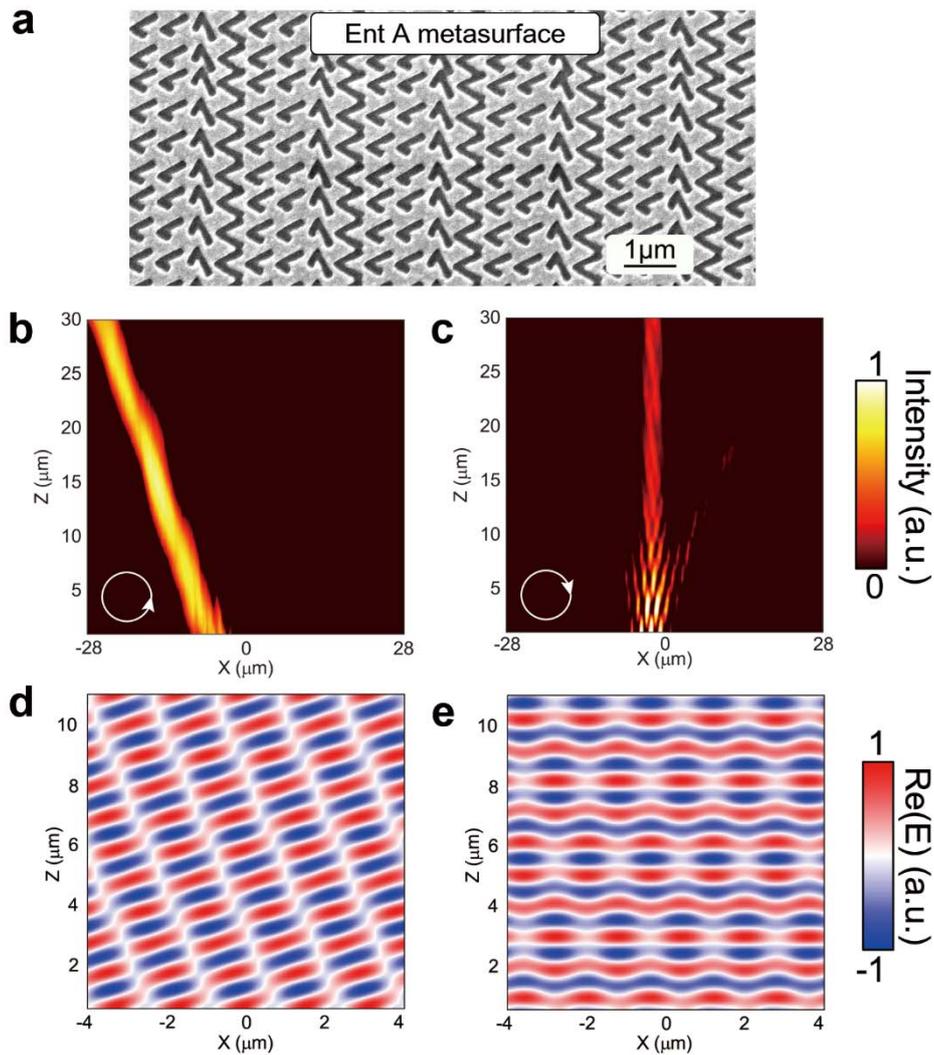

**Figure 4.** (a) SEM image of the fabricated chiral metasurfaces. (b) and (c) is measured transmitted distributions for RCP-to-LCP and LCP-to-RCP conversions, respectively. (d) and (e) are corresponding simulated results for (b) and (c).

With the ability to independently control the LCP and RCP light, the proposed spin-decoupled phase modulation has shown its novel potential in applications, in addition to the conventional method that combines propagation- and geometric-phase. To boost the functionality of our devices, the geometric phase was introduced to some of the enantiomers. As shown in **Figure 5a**, we designed a two-dimensional chiral metasurface that can simultaneously deflect light into horizontal/ vertical directions for RCP/ LCP incidences. The phase arrangements for RCP and LCP incidences are marked with the red and green rectangles, respectively. The phase interval equals 90 degrees, corresponding to a designed refraction angle 36.2 degrees. To detect the refraction along with the horizontal and vertical directions simultaneously, we collected the transmission without polarization filters. The results are shown in **Figure 5b/5c** with normal incident light whose polarization is RCP/LCP, the intensity distributions were recorded at $z = 30$ μm plane. The central bright spot in **Figure 5b/5c** represents directly transmitted light with a refraction angle of 0 degree, while the weak spots on the sides are deflected light with a non-zero refraction angle. The 20.5 μm displacement of the two spots along the horizontal direction in **Figure 5b** represents a refraction angle of 34.3 degrees for the lateral spots, which is a little smaller than the refraction angle of our theoretical evaluation (36.2 degrees). Meanwhile, the 22.2 μm displacement along the vertical direction in **Figure 5c** represents a refraction angle of 36.3 degrees, which is in good agreement with our theoretically expected refraction angle. The chiral metasurface shows two-dimensional beam steering for both RCP and LCP incidence, indicating that our enantiomers can achieve arbitrary functions of RCP and LCP independently. For example, independent holograms (vortex beam generations) for RCP and LCP incidences, etc.

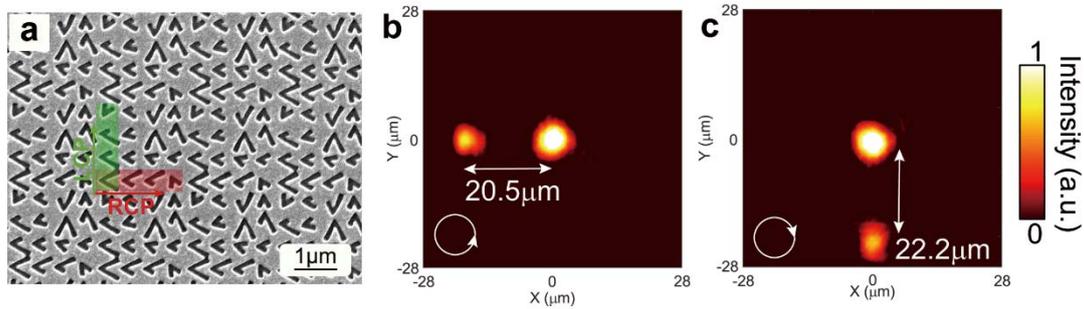

**Figure 5.** (a) SEM image of the two-dimensional chiral metasurface. Some enantiomers have an in-plane rotation. The measured intensity distribution at $z=30$μm plane for (b) RCP and (c) LCP incidences, respectively.

In conclusion, we proposed a new method to manipulate the spin phases of scattering light, which can decouple the phase responses of RCP and LCP component of light. The key feature of this phase modulation originates from chiral nanostructures that support independent control of distinguishable modes due to near-field resonant interference. To demonstrate the proposed spin-decoupled phase modulation, we first fabricated an Ent A metasurface, allowing anomalous refraction only for the transmitted cross-polarization component of the RCP incidence. Then a two-dimensional chiral metasurface was fabricated with anomalous refraction along with the horizontal/vertical directions of its cross-polarization transmission when irradiated with RCP/LCP light, respectively. The measured results of our fabricated chiral metasurfaces, simulated results, and theoretical results can match each other, verifying the capability of spin-decoupled phase modulation by our proposed enantiomers. The results presented here have significant implications in the field of metasurface to expand the horizon of phase engineering, which can pave the way for metasurfaces in spin-based devices, including polarization imaging, optical manipulation, holograms, and sensing. It is worth mentioning that the decoupled phase response presented here is not limited to our enantiomers, it still works fine in other chiral structures with proper design.

**Acknowledgment:**

National Natural Science Foundation of China (NSFC) (12174052)

# Supporting information

1. The transmission amplitude of enantiomers.

   Firstly, we consider the amplitudes for enantiomer A. The simulated amplitude for LCP and RCP incidences is shown in **figure S1**, where two dashed lines are co-polarization of LCP/RCP incidences. They are almost the same as each other. The blue and orange solid lines have some ignorable differences. However, due to their small value of difference, the circular dichroism (CD) in our enantiomers is very small. In the following parts, we can ignore the influence of CD in our chiral metasurfaces.

   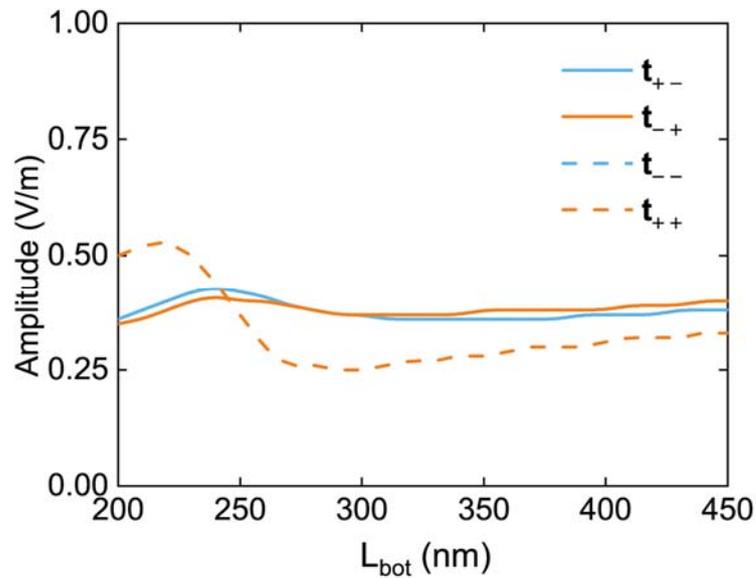

   **Figure S1.** The amplitude coefficients for transmitted lights, the solid blue line ($t_{+-}$) is the cross-polarization of LCP incidence, solid orange line ($t_{-+}$) is the cross-polarization of RCP incidence, dashed blue line ($t_{--}$) is the co-polarization of LCP incidence and dashed orange line ($t_{++}$) is the co-polarization of RCP incidence, respectively.

2. Complementary structures of V-shaped nanoholes.

   The complementary structures of V-shaped nanoholes are considered as gold chiral V-shaped antennas (CVAs) above the gold layer, as schematically shown in **figure S2a**. The phase modulations are plotted in **figure S2b**, where we only compared the phase modulation of cross-polarization for LCP and RCP incidences (blue and red lines, respectively). The phase modulation is still decoupled, only the orange line can reach a phase modulation of almost 360 degrees with varied $L_{bot}$.

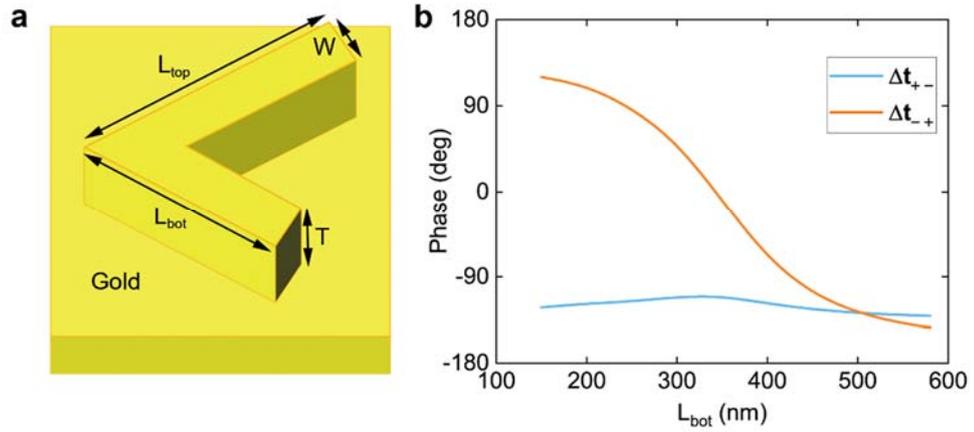

**Figure S2.** (a) schematic of CVAs, the $L_{top}$, thickness T, and width W are constant values of 500 nm, 120 nm, and 100 nm, respectively. The period is 600 nm and the CVAs are immersed in water whose refractive index is 1.33. (b) The phase for cross-polarization of LCP and RCP incidences.

3. CVAs metasurfaces

As mentioned in the main text, the full-range-covered metasurface can achieve the same two-dimensional phase modulation without any geometric phase. By using the CVAs in **figure S2**, we designed a two-dimensional chiral metasurface as schematically shown in **figure S3a**, only the bottom/top arms are changed without any additional in-plane rotation. The simulated beam steering results are shown in **figure S3b** and **S3c** for RCP and LCP incidence, respectively. The refraction angle of both incidences is 26 degrees, matching well with the theoretical value of 26.3 degrees.

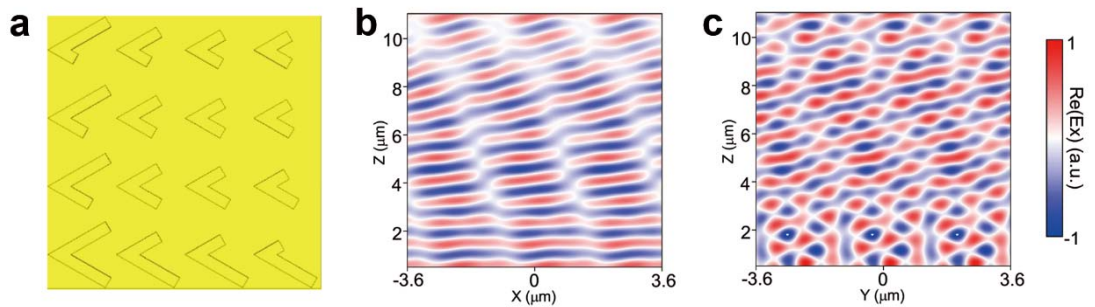

**Figure S3.** (a) Model of designed CVAs-based metasurfaces. Simulated reflected electric field distribution for (b) RCP incidence and (c) LCP incidences.

4. Fabrication process.

The designed chiral metasurfaces were fabricated by the following steps. First, a chromium film with 2-nm thickness was sputtered on a glass substrate to serve as the adhesion layer, followed by the deposition of a 40-nm-thick silver film.

Then, the 40-nm-thick silver film was etched by a focused ion beam system (FEI, FIB 200) with an ion current of 35 pA and voltage of 30 kV.

5. Optical path.

The homemade setup used to characterize chiral metasurfaces is shown in **figure S4**. A 1064-nm continuous laser was used as the source, to control the polarization of incident light, a pair of a half-wave plate (HWP) and a quarter-wave plate (QWP1) was placed in front of the laser to generate LCP/RCP light. Then, the laser passed through a beam expander consisting of two plano-convex lenses (L1 and L2, focal length 5 cm and 10 cm, respectively). After expanding its diameter, the collimated laser beam was converged by another plano-convex lens (L3, focal length 15 cm), then passed through the back focal plane of an objective lens (O1, 50×, NA=0.5). In other words, the L3 and O1 composed another beam expander. Therefore, as marked by the green dashed rectangle in figure S3, the final output light was a quasi-collimated beam with an invariant diameter of around 5 μm. In this way, the distance between the sample stage and O1 will not influence the final measured intensity distribution (the incident beam was collimated rather than a focused beam). The sample was mounted at a motorized translation stage (PT3/M-Z8, Thorlabs inc.), in our measurement, the sample movement was controlled over a total distance of 30 μm with an interval distance of 1 μm (30 frames for a distance of 30 μm), the movement was marked by a gray arrow in figure S3. An objective lens (O2, 50×, NA=0.8) was used to collect transmitted light, then the collected light was directed into a quarter-wave plate (QWP2) to degrade LCP/RCP light into linear polarized light. Whereafter, the linear polarized light was filtrated by a Glan-Taylor Calcite polarizer (GP1), which allows us to separate the intensity of cross- and co-polarization of LCP/RCP incidences. Finally, the intensity was recorded by a Camera, focused by a plano-convex lens (L4).

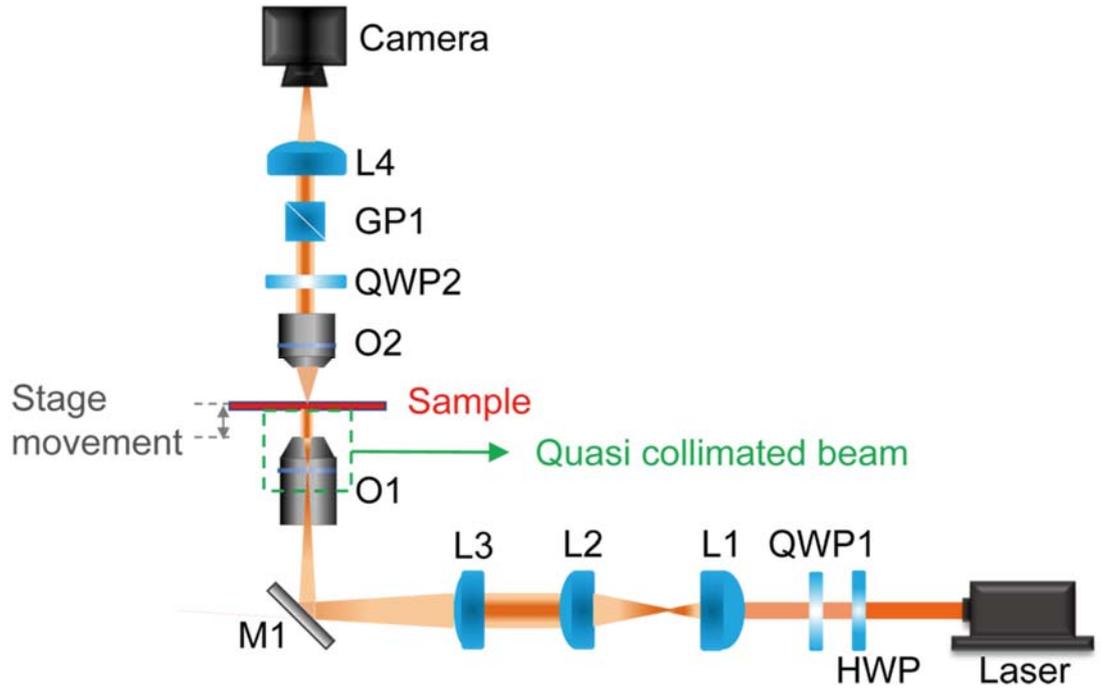

**Figure S4.** Schematic of the optical path.

6. Measured co-polarization of Ent A metasurface.

   The transmitted light of co-polarization was plotted in **figure S5a** and **S5b**. **Figure S5a** is the intensity distribution of RCP incidence (see white circle arrow for polarization, counterclockwise is RCP, clockwise is LCP). The main portion of transmitted light for either RCP or LCP has a refraction angle of 0 degree. But a little part of light has a refraction angle of 35.75 degrees, this may be due to sample imperfection and measurement error.

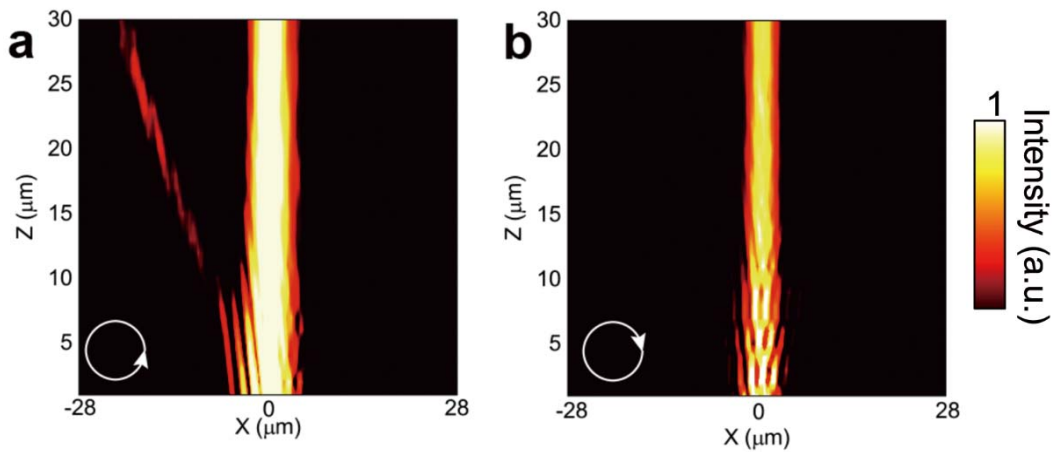

**Figure S5.** Measured transmitted light of co-polarization for (a) RCP incidence and (b) LCP incidence.

7. Ent B-based chiral metasurface.

The chiral metasurfaces composed of Ent B (the enantiomer metasurfaces of **figure 4b**) were also fabricated and measured, its scanning electron microscope (SEM) image is shown in **figure S6**. The beam deflection will only be observed for LCP incidence, as shown in **figure S7b**. Furthermore, the measured refraction angle is 34.6 degrees, a bit smaller than the theoretical refraction angle of 36.2 degrees. As for RCP incidence, not only the cross-polarization (**figure S7a**) but also the co-polarization (**figure S7c**) of transmitted light has a refraction angle of 0 degree, manifesting zero phase-gradient modulation for RCP incidence.

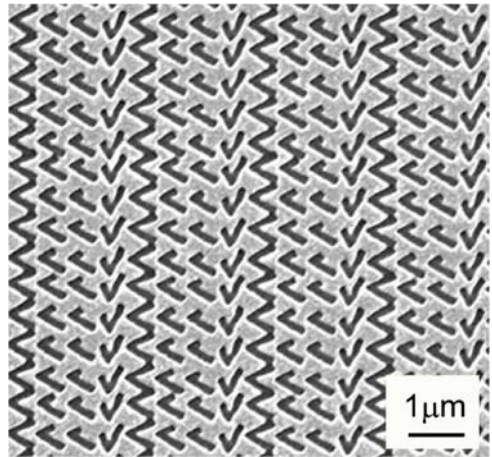

**Figure S6.** SEM image of Ent B-based metasurfaces.

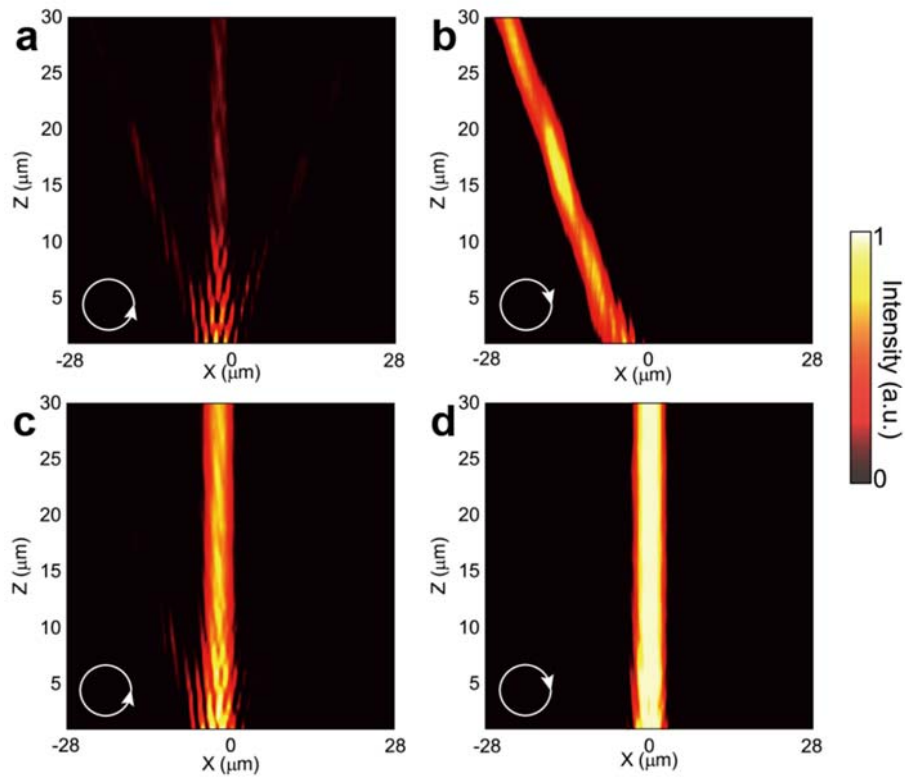

**Figure S7.** Measured transmitted intensity of Ent B metasurfaces (a) RCP-to-LCP conversion, (b) LCP-to-RCP conversion. The co-polarization for (c) RCP incidence, and (d) LCP incidence, respectively.

8. Sym-based metasurfaces.

   In **figure S8**, the symmetric structure-based (Sym-based) metasurfaces are considered for comparison with enantiomers-based metasurfaces (**figure 4** and **figure S7**). We know Syms impart the same phase modulation for LCP and RCP incidences, therefore, the beam deflection would happen for both LCP and RCP incidences. For example, the measured transmitted light in **figure S9a** and **figure S9b** have the same refraction angle, which measured value is 34.7 degrees. While for co-polarization of RCP and LCP incidences, the **figure S9c** and **figure S9d** show no obvious refraction. The phenomena above can be well explained by the typical propagation phase. In another word, our proposed chiral metasurfaces based on enantiomers break the definition of propagation phase, providing another way to design the phase modulation for metasurfaces.

   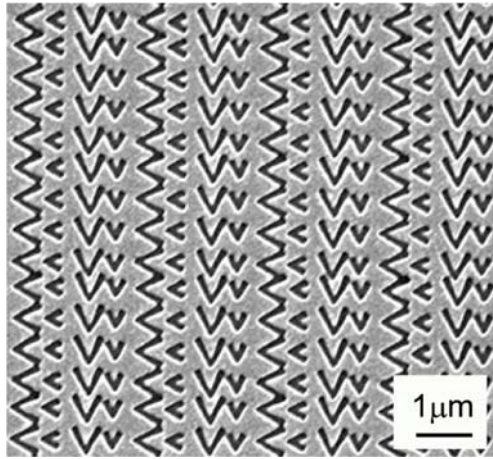

   **Figure S8.** SEM image of symmetric metasurfaces.

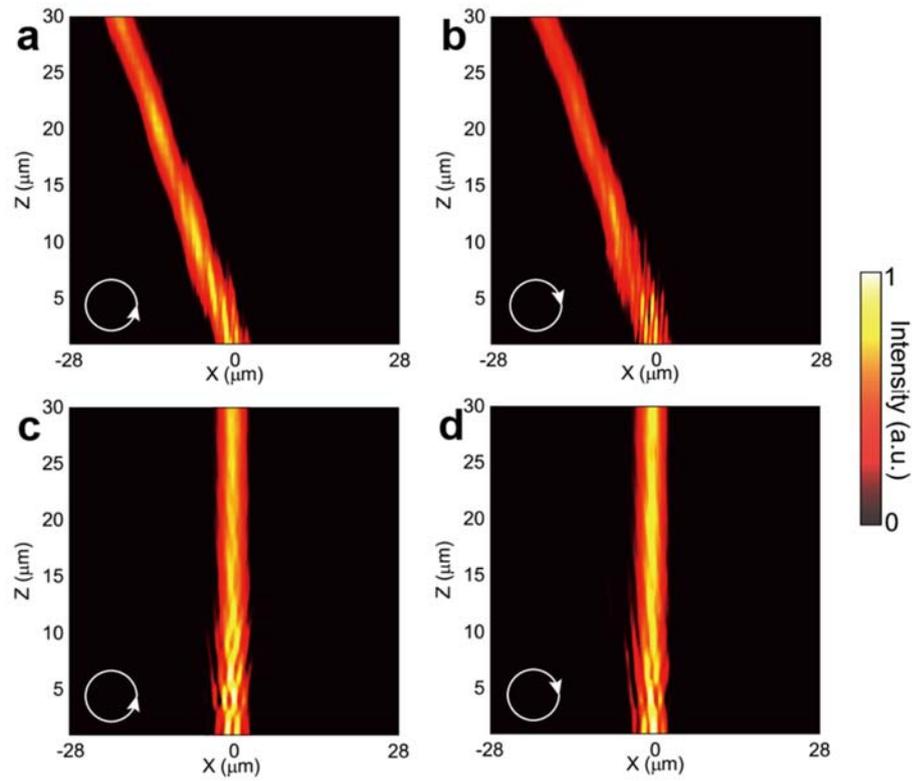

**Figure S9.** Measured transmitted intensity of symmetric metasurfaces (a) RCP-to-LCP conversion, (b) LCP-to-RCP conversion. The co-polarization for (c) RCP incidence, and (d) LCP incidence, respectively.